\documentclass[aps,prd,preprintnumbers,showpacs]{revtex4}
\usepackage[dvips]{graphicx}
\begin{document}

%
%

\preprint{Nisho-1-2011}
\title{Schwinger Mechanism with Energy Dissipation in Glasma}
\author{Aiichi Iwazaki}
\address{International Economics and Politics, Nishogakusha University,\\ 
6-16 3-bantyo Tiyoda Tokyo 102-8336, Japan.}   
\date{May 24, 2011}
\begin{abstract}
Initial states of glasma in high energy heavy ion collisions
are longitudinal classical color electric and magnetic fields. 
Assuming finite color electric conductivity,
we show that the color electric field decays by quark pair production with the life time of the order
of $Q_s^{-1}$, i.e. the inverse of the saturation momentum. Quarks and anti-quarks created in 
the pair production
are immediately thermalized as far as their temperature $\beta^{-1}$ is lower than $Q_s$.   
Namely, a relaxation time of the quarks to be thermalized 
is much shorter than $Q_s^{-1}$ when $\beta^{-1} \ll Q_s$. 
We also show that the quarks acquire longitudinal momentum of the order of $Q_s$ by
the acceleration of the electric field.
To discuss the quark pair production,
we use chiral anomaly which has been shown to be 
very powerful tool in the presence of strong magnetic field.    
\end{abstract}
\hspace*{0.3cm}
\hspace*{1cm}

\maketitle


\section{Introduction}
Initial states of color gauge fields (glasma) produced immediately after high energy heavy-ion collisions
have recently received much attention. The gauge fields are longitudinal classical color electric and magnetic fields;
their field strengths are given by the square of the saturation momentum $Q_s$.
The presence of such classical gauge fields 
has been discussed on the basis of a fairly reliable
effective theory of QCD at high energies, that is, a model of color glass condensate (CGC)\cite{gl,cgc}.
It is expected that
the decay of the glasma leads to thermalized quark gluon plasma (QGP). 
 
The glasma is homogeneous in the longitudinal
direction and inhomogeneous in the transverse directions. Hence,
we may view that it forms electric and magnetic flux tubes extending in the longitudinal direction. 
In the previous papers\cite{iwa,itakura,hii} we have shown a possibility that 
the famous Nielsen-Olesen instability\cite{nielsen} makes the color magnetic field $B$
decay.  The possibility has been partially confirmed by
the comparison between our results and numerical simulations\cite{venugopalan,berges}. 
Such decay is a first step toward the generation of QGP.

On the other hand we have also discussed\cite{iwazaki1,iwazaki2,old} the decay of the color electric field; the decay
is caused by Schwinger mechanism\cite{schwinger}, that is, the pair production of quarks and anti-quarks.
The mechanism has been extensively explored\cite{tanji} since the discovery of Klein paradox.
Among them, the pair production in the expanding glasma has been discussed\cite{lap}. 
A new feature in the glasma is that it is composed of not only electric field
but also magnetic field parallel to the electric field. Such a feature has also been
explored. In particular, recently there are studies of the back reaction of the particles on the electric field
under the presence of the magnetic field\cite{iwazaki2,tanji,suga}.  
The back reaction is essentially important for the decay of the electric field.

Our originality\cite{iwazaki2} for the investigation of the decay
is to use chiral anomaly. As is well known, the anomaly is effective when collinear 
magnetic and electric fields are present. This is the very situation in the glasma. 
When we use the chiral anomaly, we can discuss Schwinger mechanism without detail calculations\cite{tanji,lap,suga}
of wave functions but simply by solving classical anomaly equation and
Maxwell equations. 
In particular, when the strong magnetic field is present,
the anomaly is much simplified
because the quarks are effectively massless and only relevant states are
ones in the lowest Landau level. 
( Both $\sqrt{gE}\sim Q_s$ and $\sqrt{gB}\sim Q_s$ in the glasma are much larger than mass of quarks. )
 Since the motions of the quarks in transverse directions are frozen,
only possible motion is along the longitudinal direction. Thus, the anomaly equation
is reduced to the one in two dimensional space-time. With the simplification,
we can find important quantities 
even in complicated realistic situations
for which the investigations have not yet performed.  
Actually, we have discussed the decay of axial symmetric electric flux tube by taking account of 
the azimuthal magnetic field around the tube. The field is generated by the current carried 
by the pair created quarks and anti-quarks.
Although the electric field loses its energy by the pair creation
and the generation of the azimuthal magnetic field, it never decays. 
The field oscillating with time propagates to the infinity in the axial symmetric way\cite{iwazaki2}. This is because
the quarks are free particles and there is no energy dissipation. 
( In the case of homogeneous electric field, the field simply oscillates with time. )

In this paper we wish to discuss the decay of the color electric field $E$
by taking account of energy dissipation in heat baths.
The dissipation arises due to the presence of finite electric conductivity. Namely,
the pair production generates color electric current $J$,
which dissipates its energy owing to
the interaction between the quarks and the surrounding;
the surrounding is composed of quarks and gluons.
Actually, we assume Ohm law $J=\sigma E $ with electric conductivity $\sigma$. 
The conductivity is calculated by using Boltzmann equation in the relaxation time approximation.
In the approximation a relaxation time is obtained by calculating electron's scattering
rates. Then,
we can show that the quarks are thermalized immediately after their production as far as their temperature 
is much smaller than $Q_s\sim \sqrt{eE}$ ;
the relaxation time of a slightly deformed momentum distribution of quarks becoming the
equilibrium Fermi distribution is much shorter than the life time $\sim Q_s^{-1}$ of the field.
As numerical calculations have shown\cite{tanji}, the longitudinal momentum
distribution of the free particles produced in the vacuum is almost equal to the equilibrium one, 
that is Fermi distribution at zero temperature. Thus, even in non zero but low temperature,
the momentum distribution is nearly equal to
the equilibrium one. Our relaxation time approximation in Boltzmann equation may holds in such a situation. 
Therefore, owing to the energy dissipation by the scattering between electrons and positrons, 
the electric field decays and never oscillates.



For simplicity, we examine homogeneous electric and magnetic fields of
U(1) gauge theory instead of SU(3) gauge theory.
Thus, we use terminology of electron or positron instead of
quark or anti-quark. The generalization of our results to the case of SU(3) gauge theory is
straightforward done simply by assuming maximal Abelian group of SU(3)\cite{tanji}.
We assume that both the electric and magnetic fields are much larger than
the square of the electron mass. Thus, they are taken to be massless.

In the next section we explain how the chiral anomaly is useful for the
discussion of Schwinger mechanism.
We apply the anomaly to the discussion of the pair production
with energy dissipation in the section $(3)$.
In the discussion we use the conductivity of electron-positron gas
under strong magnetic field. The detail calculation of it is shown in
the appendix.

\section{Chiral anomaly and pair production}

We first make a brief review of the utility of the chiral anomaly.
As we will show below, only by simple calculations 
we can obtain most of important results of Schwinger mechanism, which was
derived previously by detail calculations.
 
Under the homogenous magnetic field $B$, massless fermions with charge $e>0$ occupy the states
characterized by Landau level and momentum $p$ parallel to the magnetic field.
Their energies are given by 

\begin{equation}
E_N=\sqrt{p^2+2NeB} \quad \mbox{(parallel)} \quad \mbox{and} 
\quad E_N=\sqrt{p^2+eB(2N+1)} \quad \mbox{(antiparallel)},
\end{equation}
where integer $N$ denotes Landau level.
The term of ``parallel" (``anti-parallel") implies magnetic moment parallel (anti-parallel) to $\vec{B}$. 
The magnetic moment of electrons (positrons) is antiparallel (parallel) to their spin.
Thus,
electrons (positrons) with spin anti-parallel (parallel) to $\vec{B}$ can have zero energy states
in the lowest Landau level; their energy spectrum is given by $E_{N=0}=|p|\ge 0$. 
On the other hand, the other states cannot be zero energy states.
Hence only relevant states for the pair production are states with energy $E_{N=0}=|p|$ 
in the limit of strong magnetic field.  Hereafter, we consider only such states.

The equation of the chiral anomaly for the states
leads to

\begin{equation}
\label{chiral}
 \partial_t (n_R-n_L)=\frac{e^2}{2\pi^2}\vec{E}(t)\cdot\vec{B}
\end{equation}
where $n_R$ ( $n_L$ ) denotes number density of right ( left ) handed chiral fermions; 
$n_{R,L}\equiv \langle\bar{\Psi}\gamma_0(1\pm \gamma_5)\Psi\rangle/2$
in which the expectation value is taken by using the states in the lowest Landau level.
We have used spatial uniformness of the chiral current $\vec{j}_5$, that is, $\rm{div}\vec{j}_5=0$.

We note that electrons move to the direction anti-parallel to $\vec{E}$ while
positrons move to the direction parallel to $\vec{E}$ after their pair production.
Therefore, both the electrons and positrons have right handed helicity
when $\vec{E}$ being parallel to $\vec{B}$. 
( Right or left handed helicity means the state with momentum parallel or antiparallel to spin, respectively.)
Since chirality is equivalent to helicity of massless fermions,
$n_R=2n$ and $n_L=0$ where $n$ is
the number density of electrons.
( When $\vec{E}$ is anti-parallel to $\vec{B}$, $n_R=0$ and $n_L=2n$. )
Therefore, the anomaly equation describes the pair production rate of electron and positrons under $E$ and $B$,

\begin{equation}
\label{ch}
\partial_t n=\frac{e^2}{4\pi^2}E(t)B \quad \mbox{for}\quad E(t)>0 \quad \mbox{and}\quad
\partial_t n=-\frac{e^2}{4\pi^2}E(t)B \quad \mbox{for}\quad E(t)<0
\end{equation}

For example, we suppose that the electric field $E(t)>0$ is switched on at $t=0$. Then,
solving this equation with initial condition $n(t=0)=0$, 
we can obtain the number density of electrons\cite{tanji}; $n(t)=\int^{t}_0\frac{e^2}{4\pi^2}E(t')Bdt'$.
When $E$ is constant, $n$ increases linearly with time.
In this estimation,
the back reaction of the produced particles 
to the electric field 
is not taken into account.

When we take account of the back reaction, 
we can find that the electric field oscillates with time. 
The back reaction is that the energy of electric field decreases owing to the acceleration of
the charged particles. 
It can be included by solving a Maxwell equation, $\partial_t E=-J$
along with the anomaly equation. To solve the equations,
we need to express the electric current $J$ in terms of $n$.
This can be done by considering the energy conservation.
Namely, the total energy of the electric field and the charged particles $\epsilon$ is conserved,

\begin{equation}
\label{4}
\partial_t \Bigl(\frac{E(t)^2}{2}+2\epsilon(t)\Bigr)=E\partial_tE+\partial_t n(t)p_F+n(t)eE=E(-J+2en(t))=0,
\end{equation}
where the energy density of electrons or positrons is given 
such that $\epsilon(t)=n(t)p_F(t)/2$
with Fermi momentum $p_F(t)=\int_0^t eE(t')dt'=n(t)4\pi^2/eB$ for $E>0$. 
Namely, we assume the energy of free electrons and positrons.
Their momentum distributions are supposed to be $\tilde{n}_0\propto \theta(p_F(t)-p)\theta(p)$ for positrons or
$\tilde{n}_0\propto \theta(p_F(t)+p)\theta(-p)$ for electrons. 
We have used the formula $\partial_tn(t)p_F=\partial_tn(t) n(t)4\pi^2/eB=en(t)E$ in eq(\ref{4}).
Thus, from the equation we obtain the electric current $J=2en$ for $E>0$.  
Obviously, there is no energy dissipation.

Using the Maxwell equation and the anomaly equation we derive the equation of the number density,

\begin{equation}
\label{nvac}
\partial_t^2n(t)+\frac{e^3B}{2\pi^2}n(t)=0.
\end{equation}

We can see that after the switch on of $E$, $E$ gradually decreases with time,
while $n$ increases. Eventually, these quantities oscillate with the frequency
$e\sqrt{eB/2\pi^2}$ ( plasma oscillation ) as shown in Fig.1.
The oscillation can be understood in the following.

\begin{figure}[htb]
    \centering
    \includegraphics*[width=65mm]{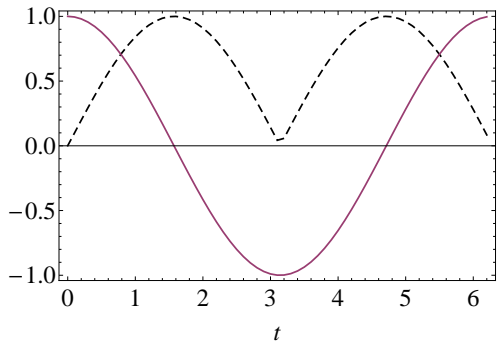}
    \caption{number density $n$ (dash)
and electric field (line) with arbitrary scale}
    \label{l2ea4-f1}
\end{figure}

When $E$ ( $>0$ ) parallel to $B$ is switched on at $t=0$ with $n(t=0)=0$, the number $n$ of the particles increases while
$E$ decreases. When $E$ vanishes, $n$ takes the maximum. After that
$E$ changes its sign, namely, $E$ becomes negative ( anti-parallel to $B$.)  
Then, the particles are forced to move
to the direction inverse to the one in the case of $E>0$.
Thus, the pair annihilation occurs and $n$ begins to decrease. 
The energy of the particles is transmitted to the energy of the electric field. 
Actually, according to the anomaly
equation,  $|E|$ becomes large with time. When $n$ vanishes, $|E|$ takes the maximum ( $E=-E(t=0)$ ).
Subsequently, the pair creation occurs and $n$ increases according to the anomaly
equation(\ref{ch}) but with $E<0$.
This is a plasma oscillation shown in the figure.
The energy of the electric field is transmitted to the particles and then the energy of the particles
goes back to the electric field.
The plasma oscillation is a natural consequence of the absence of the dissipation.

We should make a comment on the scales involved in the discussion.
The above result ( plasma oscillation ) holds in the limit of $ eB \gg eE $ where
only the states in the lowest Landau level are relevant. On the other hand when $eB \ll eE $,
states in higher Landau levels become important since the energy $|p_F(t)|=|\int_0^t dt' eE|$ of electrons can be
much larger than $\sqrt{eB}$.
In the glasma we have  both $eB $ and $ eE\sim Q_s^2 \gg m_{\rm{electron}}^2$ ( square of electron mass ). But 
states in higher Landau levels are still not important according to the following reason.
When some of electrons in the lowest Landau level are accelerated and their energies reach the energy 
$|p_F|=\int_0^{Q_s^{-1}} dt' eE \simeq eE Q_s^{-1}\sim \sqrt{gB}$,
they may make a transition to a state in a higher Landau level with energy $=\sqrt{gB}$.
Such electrons are the ones produced in the very early stage of the pair production.
The number of the electrons is still a fraction. 
Most of electrons whose energies are less than $\sqrt{eB}$, occupy the lowest Landau level.
It has recently been shown\cite{suga} that the decay of the electric field is mainly caused by particles 
in the lowest Landau level. 
Thus, our result is 
approximately correct.

\vspace{0.3cm}

In this way, we obtain the number density of electrons
simply by using the chiral anomaly. It is important to note that
the anomaly equation involves all of quantum effects. Thus, by simply solving the classical anomaly
equation, we can obtain the quantities resulting from
the purely quantum effect, i.e. Schwinger mechanism. 
In the standard manipulation\cite{tanji} of Schwinger mechanism, we need to find
wave functions of electrons under the electric and magnetic fields
in order to obtain such quantities. When the time dependence of the electric field is complicated,
it is very difficult to obtain the wave functions. On the other hand,
the anomaly equation is a very simple and powerful tool
for investigating Schwinger mechanism.

\vspace{0.3cm}
\section{Pair production with energy dissipation}

Up to now, we have not taken into account energy dissipation of electrons and positrons produced by
the electric field.
We proceed to analyse the Schwinger mechanism in heat bath, especially,  
the case with energy dissipation.
We assume that the particles are immediately thermalized after their production.
( Later we see that relaxation time, within which slightly deformed particle distribution becomes
equilibrium distribution, is much shorter than the typical scale $Q_s^{-1}\simeq \sqrt{1/eE}$ 
as far as the system is in low temperature. As numerical calculations have shown\cite{tanji}, the momentum
distribution of the produced particles is nearly equal to the Fermi distribution at zero temperature. )
Thus, the number density $n$ is given by

\begin{equation}
n(t)=a_0\int^{+\infty}_{0} dp\frac{1}{\exp{(\beta(t)(|p|-\mu(t)))}+1},
\end{equation}
with $a_0=eB/2\pi^2$,
where $\beta=1/T$ is the inverse of temperature $T$ and $\mu$ is chemical potential of electrons or positrons.

The number density $n$ depends on two parameters, $\beta$ and $\mu$, which may be functions of time.
Hereafter, we consider only the case that the chemical potential $\mu$ vanishes due to
non-conservation of electron number. It is caused by
frequent pair annihilation or creation of electrons and positrons after their production.
Since initial states have no lepton number, the chemical potential vanishes.
( This corresponds to the situation in the glasma where the pair annihilation 
or creation of quarks and anti-quarks frequently take place
 after the pair creation by Schwinger mechanism. )  
Thus, the number density is only a function of the temperature and is given by
$n=a_0\beta(t)^{-1}\log2$. 
Similarly, the energy density is given by

\begin{equation}
\epsilon(t)=a_0\int^{+\infty}_{0} dp\frac{p}{\exp{\bigl(\beta(t)p\bigr)}+1}=c_0a_0\beta(t)^{-2}
\end{equation}
with $c_0=\int_0^{\infty}x\,dx/(\exp(x)+1)=\pi^2/12$. 
Since the temperature $\beta^{-1}$ represents an average energy of the particles,
it is reasonable that we have
$\epsilon \sim n\beta^{-1}$; the energy density is equal to the number 
density times the average energy of a particle.

The electric field produces electrons and positrons 
and generates their electric currents.
When the electric current flows,
the current dissipates its energy due to the finite
electric conductivity of these particles.
Thus, the electric field
dissipates its energy.
Here we may assume that the electric current $J$ 
satisfies the Ohm law $J=\sigma(\beta) E$
where $\sigma$ denotes electric conductivity. 
Obviously, the Ohm law represents the energy dissipation of the electric field
$\partial_t (E^2/2)=E\partial_t E=-EJ=-\sigma  E^2<0$.

\vspace{0.2cm}

As we will show in the appendix, the electric conductivity $\sigma$ of the charged particles in the lowest Landau level
is given by

\begin{equation}
\label{co}
\sigma=\frac{e^2\beta eB}{4\pi^2} \int_{-\infty}^{+\infty} dp \tau(p) f_0(p)(1-f_0(p)),
\end{equation}
where $f_0\equiv 1/\bigl(\exp(|p|\beta)+1\bigr)$ 
denotes the momentum distribution of the charged particles. The conductivity
depends on the relaxation time $\tau(p)$; the time needed for a momentum distribution
to become the equilibrium one $f_0(p)$. The relaxation is realized by  
scattering of the charged particles so that it can be expressed by their
scattering amplitudes. The detail calculation shows that $\tau \to (eB\beta)^{-1}$ in the limit 
of low temperature $\beta^{-1}\ll \sqrt{gB}$.
( This fact makes valid our assumption of the immediate thermalization in the low temperature
where the relaxation time is small. )
Thus, the conductivity depends on the temperature 
such as $\sigma(\beta)\simeq b_0\beta^{-1}(t)$ in the limit
( $b_0$ is a numerical constant ).

Thus,  
using the Maxwell equation $\partial_t E=-J=-\sigma E$ and 
the anomaly equation $\partial_t n=\gamma E$ ( $\gamma \equiv e^2B/4\pi^2$ ),
we find that

\begin{equation}
\partial_t^2 \log E(t)=-\partial_t \sigma=-b_0\partial_t \Bigl(\frac{n}{a_0\log 2}\Bigr)
=-b_0\Bigl(\frac{\gamma E}{a_0\log 2}\Bigr)=
-\frac{eb_0}{2\log 2}E(t).
\end{equation} 

Setting $X\equiv \log \bigl(E(t)/E(t=0)\bigr)$, we obtain the equation of the electric field

\begin{equation}
\label{X}
\partial_t^2 X(t)+\frac{b_0eE(t=0)}{2\log 2}\exp \bigl( X(t)\bigr )=0
\end{equation} 
with initial conditions $X(t=0)=0$ and $\partial_t X(t=0)=0$.
 The number density $n(t)$ can be obtained by integrating the solution $E(t)=E(t=0)\exp( X(t) )$
such that $n(t)=\gamma \int_0^t dt'E(t')$. 
 
The equation (\ref{X}) 	describes a motion of a classical particle with the potential energy $\sim \exp \bigl( X \bigr )$;
$X$ denotes the coordinate of the particle. Starting at $X=0$,  the particle goes down 
to $-\infty $ along the slope of the potential. It means that $E\propto \exp \bigl(X \bigr) \to 0$.
Namely, the electric field vanishes with the life time of the typical scale $1/\sqrt{eE(t=0)}\sim 1/Q_s$.
( The life time of the electric field is 
sufficiently short to be consistent with phenomenology\cite{hirano} of QGP. )
Since $n(t)=\gamma \int_0^t dt'E(t')=a_0T(t)\log2$, 
the number density $n$ or the temperature $T$ monotonically increases with time
and reaches its maximum.
This is contrasted with the case of no dissipation in which 
both $E$ and $n$ oscillate.  
These behaviors are shown in the Fig.2.

\begin{figure}[htb]
 \centering
 \includegraphics*[width=65mm]{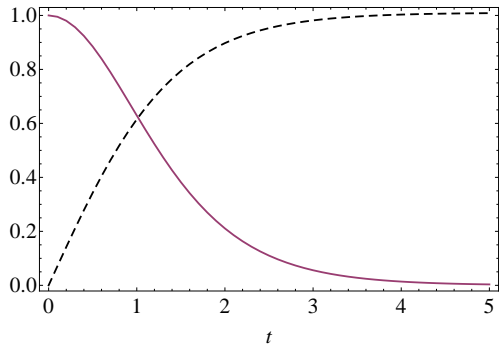}
 \caption{number density $n$ (dash)
and electric field (line) with arbitrary scale}
    \label{l2ea4-f1}
\end{figure}

We can roughly estimate the attainable number density $n$, temperature $T$ and average energy of the particles
$\sim \epsilon/n$
in the pair creation,

\begin{eqnarray}
n&=&\frac{e^2B}{4\pi^2}\int_0^{\infty}dt'E(t')\simeq \frac{e^2B}{4\pi^2}\frac{E(t=0)}{\sqrt{eE(t=0)}}
\sim 0.01eB\times \sqrt{eE(t=0)} \quad \nonumber \\ 
T&\sim& \sqrt{eE(t=0)} \quad \mbox{and} \quad  \frac{\epsilon}{n}\sim T 
\end{eqnarray}
where we set $e^2\simeq 1$ which corresponds to gauge coupling constant $g^2\simeq 1$ in QCD
with the energies at RICH or LHC.

We should make a comment.
The result has been derived in the limit of the low temperature.
The relaxation time $\tau\propto \beta^{-1} $ is sufficiently short in the limit 
so that we can use the equilibrium momentum distribution.
The temperature increases with time from zero after the switch-on of the electric field.
Therefore, the result is reliable in the early stage of the pair creation. However, 
when it becomes comparable to $\sqrt{gB}$, the assumption does not hold.
Then, we cannot use the formula $\sigma \sim \beta^{-1}$, $\tau\propto \beta^{-1} $ 
and the equilibrium momentum distribution.
Hence, the result of the saturation in the number density at the late stage of the evolution cannot be
reliable.  
But, as the temperature increases and becomes comparable to $\sqrt{gB}$,
the conductivity is almost constant in time,
Then, we find that $E=E_0\exp(-\sigma t)$ from the equation $\partial_tE= -\sigma E$ 
and that $n=n_0-\gamma \sigma E_0\exp(-\sigma t)$ from the anomaly equation.
The number density is saturated and never oscillate. ( To obtain this result we do not use
the equilibrium momentum distribution. Namely, the assumption of the particles being immediately thermalized is not
necessary. ) 
Actually, when the temperature or the average energy of the particles is comparable with $\sqrt{gB}$,
the electric conductivity may depends weakly on the time 
because only dimensional quantity in $\sigma$ is $\sqrt{gB}$ or the average energy.
Consequently, we find that the number density ( temperature ) gradually increases in the early stage 
and is saturated at the late stage.
It never oscillates unlike the case with no energy dissipation.

To put it another way,
in the early stage when the pair creation starts, the particles are thermalized soon because
the relaxation time is short. Thus, the temperature can be defined. The temperature or 
the average energy of the particles is much less than $\sqrt{eE}\sim Q_s$.
But in the stage when relaxation time is not sufficiently short for the particles to be
thermalized soon, the temperature cannot be defined. In the stage all of the typical energy scales
are of the order of $Q_s$. Thus, the electric conductivity is also of the order of $Q_s$.
This leads to the above result of the saturation $n=n_0-\gamma \sigma E_0\exp(-\sigma t)$. 
The stage is not early stage but may be
late stage in the pair creation. In this way
we understand that the results shown in Fig.2 are reliable.

\section{discussion and summary}

We have shown that the color electric field decays into the quarks and anti-quarks under
the color magnetic field. They occupy only the states in the lowest Landau level.
We have also discussed that the average energy of the particles is of the order of 
the saturation momentum $Q_s$. Namely, the Fermi momentum, which is the longitudinal momentum,
is given by $Q_s$. The particles obtain the longitudinal momentum with the acceleration by the electric field. 
We remind you that the glasma has only transverse momentum of the order of $Q_s$ and has no longitudinal momentum
owing to the homogeneity in the direction. Thus, initially there is anisotropy in momentum distribution.
But owing to the decay of the electric field, the particles have longitudinal momenta
of the order of $Q_s$.
Thus, the longitudinal and transverse momenta become almost identical to 
each other.
This is a step toward the isotropy of the momenta or thermalization of the quarks.

The life time of the electric field has been found to be of the order of $Q_s^{-1}$,
while that of the magnetic field by a decay mechanism of Nielsen Olesen instability is much longer\cite{venugopalan} than $Q_s^{-1}$.
We may speculate that the magnetic field also decay rapidly due to Schwinger mechanism.
This is because azimuthal electric field is generated by
the expansion of the longitudinal magnetic flux tube and this electric field decays
by Schwinger mechanism. The expansion in the transverse direction of the magnetic flux tube is simply a result
of Maxwell equations\cite{iwa,itakura}. Similarly, azimuthal magnetic field is also generated by the expansion
of the longitudinal electric flux tube. Thus, both azimuthal magnetic and electric fields are
generated around the longitudinal flux tubes.  Then, the azimuthal electric field decays in the way as we have shown.
The field strengths of these longitudinal and transverse ( azimuthal ) fields
have been shown\cite{lappi} to become the same order of magnitudes at about the time $Q_s^{-1}$ after 
the high energy heavy ion collisions.
In this way, the longitudinal magnetic flux may decay rapidly through the decay of the azimuthal electric field
generated by the expansion of the magnetic flux.
Such a rapid decay of the magnetic field as well as the electric field is favorable for establishing
early thermalization\cite{hirano} of QGP.

To summarize, we have shown how the electric field decays by Schwinger mechanism
in the presence of the energy dissipation. Namely, the electric current carried by 
electrons and positrons dissipate its energy owing to the Ohm law, which
results in the increase of
the temperature with time.
Our result holds only in the limit of low temperature, in which
the relaxation time has been shown to be much short; 
the thermalization of produced particles can be realized immediately after their production.
However, we have also discussed that our result is approximately correct
even when the temperature or the average energy of the particles is of the order of $Q_s$.


\vspace{0.2cm}

\appendix
\section{electric conductivity {\Large $\sigma $}}

We show that the electric conductivity behaves as $\sigma \propto \beta^{-1}$
in the limit of the low temperature $gB\beta^2 \to \infty$. We calculate $\sigma$ in the relaxation time approximation
of Boltzmann equation,

\begin{equation}
\dot{p}_i\partial_{p_i}f=\dot{p}_z\partial_{p_z}f=eE\partial_p f=-\frac{f-f_0}{\tau}
\end{equation}
with $p=p_z$,
where we have used classical equation of motion, 
$\dot{\vec{p}}=e\vec{E}+e\vec{p}/|p|\times \vec{B}$.
$\tau$ denotes relaxation time and $f_0$ does
Fermi momentum distribution 
with the vanishing chemical potential; $f_0=1/\bigl(\exp(|p|\beta)+1\bigr)$ . 
$f$ represents a deformed momentum distribution caused by the effect of the 
electric field.
Since the electric current $J$ is given by 

\begin{equation}
J=e\frac{a_0}{2}\int_{-\infty}^{+\infty} f(p)\frac{p}{|p|}dp=e\frac{a_0}{2}\int_{-\infty}^{+\infty} (f-f_0)\frac{p}{|p|}dp
\simeq -\frac{a_0}{2}e^2E \int_{-\infty}^{+\infty} \tau\partial_pf_0\frac{p}{|p|}dp
=\frac{a_0}{2}e^2E\beta \int_{-\infty}^{+\infty} \tau f_0(1-f_0)dp,
\end{equation}

we obtain the electric conductivity 

\begin{equation}
\sigma=\frac{e^2\beta eB}{4\pi^2} \int_{-\infty}^{+\infty} dp \tau(p) f_0(p)(1-f_0(p)).
\end{equation}

The relaxation time $\tau(p)$ is given by the inverse of the scattering rate $W(p)$
between electrons and positrons,

\begin{eqnarray}
&e^-(p)&\,\,\,+\,\,\,e^-(q)\,\,\,\to \,\,\,e^-(k)\,\,\,+\,\,\,e^-(l) 
\quad \mbox{and} \quad e^-(p)\,\,\,+\,\,\,e^+(q)\,\,\,\to \,\,\,e^-(k)\,\,\,+\,\,\,e^+(l)\\
\tau(p)&=&\frac{1}{W_{\rm{dir}}(e^-e^-)}+\frac{1}{W_{\rm{ex}}(e^-e^-)}+\frac{1}{W_{\rm{cross}}(e^-e^-)}
+\frac{1}{W_{\rm{an}}(e^+e^-)}\nonumber \\
&&+\frac{1}{W_{\rm{dir}}(e^+e^-)}+\frac{1}{W_{\rm{ex}}(e^+e^-)}+\frac{1}{W_{\rm{cross}}(e^+e^-)}
\end{eqnarray}
where we denote longitudinal momentum of each particle ( parallel to $\vec{B}$ ) as $p$, $q$, $k$,and $l$.
We use the notations like $W_{\rm{dir}}$, etc. to distinguish the contributions of
different Feymann diagrams,
which was used in the paper \cite{W}.
We also use the notations $W(e^-e^-)$ and $W(e^+e^-)$ to distinguish the contributions
of the processes $e^-e^-\to  e^-e^-$ and $e^+e^-\to  e^+e^-$.                                       

Our concern is scattering of the massless fermions occupying only the states in the lowest Landau level.
Their wave functions are given by

\begin{equation}
\label{sol}
\Psi_p=\frac{1}{\sqrt{2L_yL_z}}\Bigl(\frac{eB}{4\pi}\Bigr)^{1/4}\exp(-iE_pt+ip_2 y+ip z)\exp\Bigl(-\frac{eB}{2}(x-\frac{p_2}{eB})^2\Bigr)u(p)
\end{equation}
with $E_p=\pm |p|$ and Dirac spinor $u(p)$,

\begin{equation}
u(p)=\left(
\begin{array}{@{\,}cccc@{\,}}
1\\0\\\ p/E_p\\0 \end{array}\right)\quad .
\end{equation} 
where we use the gauge potential $A=(0,xB,0)$ of the magnetic field $\vec{B}=(0,0,B)$.
$L_y$ and $L_z$ denote spatial lengths in the direction of $y$ and $z$ axis, respectively.
The states are characterized by two momenta $p$ and $p_2$.
Using the formulae, the scattering amplitudes can be calculated in the following way.
For example, the scattering amplitude $\Pi$ of $e^-(p)\,\,\,+\,\,\,e^-(q)\,\,\,\to \,\,\,e^-(k)\,\,\,+\,\,\,e^-(l)$
associated with $W_{\rm{dir}}(e^-e^-)$ is obtained by estimating the formula,

\begin{equation}
\int d^4x d^4y J_{\mu}(x;p,k)J^{\mu}(y;q,l)\frac{i\exp(iq'(x-y))}{q'^2}\frac{d^4q'}{(2\pi)^4}
\equiv \Pi(2\pi)^3\delta(p_0+q_0-k_0-l_0)\delta(p_y+q_y-k_y-l_y)\delta(p+q-k-l)
\end{equation}
with $J_{\mu}(x;p,k)\equiv e\bar{\Psi}_p(x)\gamma_{\mu}\Psi_k(x)$ and $q'x=q'_0x^0-q'_1x^1-q'_2x^2-q'_3x^3$.
Here we also used another notations $t=x^0,x=x^1,y=x^2, z=x^3$ in eq(\ref{sol}).
Using the amplitude $\Pi$, we define the scattering rate $W_{\rm{dir}}(e^-e^-)$,

\begin{equation}
dW_{\rm{dir}}(e^-e^-)=|\Pi|^2(2\pi)^3\delta(p_0+q_0-k_0-l_0)\delta(p_y+q_y-k_y-l_y)\delta(p+q-k-l) \frac{L_y^4L_z^4dq_ydqdk_ydkdl_ydl}{(2\pi)^6}.
\end{equation}
The other scattering rates can be obtained in the similar way.
Consequently, we find the formulae of $W(p)$ which are fairly simplified
since the fermions are massless,

\begin{eqnarray}
\label{wd}
&W_{dir}(e^-e^-;p)&=\frac{e^4}{4(2\pi)^4} \int_{-\infty}^{+\infty} dqdkdldq_2dk_2dl_2\delta(p+q-k-l)\delta(p_2+q_2-k_2-l_2)
\delta(|p|+|q|-|k|-|l|) \nonumber \\
&\times & \frac{1}{(|p||q||k||l|)^2Z_{dir}^2}\exp\Bigl(-\frac{k_2^2-Z_{dir}^2}{eB}\Bigr)
\Bigl(\sqrt{\frac{\pi}{2}}\exp(-\frac{Z_{dir}|l_2|}{eB})
-\int_0^{\frac{Z_{dir}^2}{2eB}} dx \frac{\exp(-x-\frac{l_2^2Z_{dir}^2}{4xe^2B^2})}{\sqrt{2x}}\Bigr)^2 \nonumber \\
&\times &\bigl((|p||k|+pk)(|q||l|+ql)-(|k|p+|p|k)(|l|q+l|q|)\bigr)^2\,\,f_0(q)(1-f_0(k))(1-f_0(l)),
\end{eqnarray}       
where $Z_{dir}^2\equiv k_2^2+2(|p||k|-pk)$,

\begin{eqnarray}
\label{we}
&W_{ex}(e^-e^-;p)&=\frac{e^4}{4(2\pi)^4}\int_{-\infty}^{+\infty} dqdkdldq_2dk_2dl_2\delta(p+q-k-l)\delta(p_2+q_2-k_2-l_2)
\delta(|p|+|q|-|k|-|l|) \nonumber \\
&\times & \frac{1}{(|p||q||k||l|)^2Z_{ex}^2}\exp\Bigl(-\frac{l_2^2-Z_{ex}^2}{eB}\Bigr)
\Bigl(\sqrt{\frac{\pi}{2}}\exp(-\frac{Z_{ex}|k_2|}{eB})
-\int_0^{\frac{Z_{ex}^2}{2eB}} dx \frac{\exp(-x-\frac{k_2^2Z_{ex}^2}{4xe^2B^2})}{\sqrt{2x}}\Bigr)^2 \nonumber \\
&\times &\bigl((|p||l|+pl)(|q||k|+qk)-(|l|p+|p|l)(|k|q+k|q|)\bigr)^2\,\,f_0(q)(1-f_0(k))(1-f_0(l)),
\end{eqnarray}
where  $Z_{ex}^2\equiv l_2^2+2(|p||l|-pl)$,

\begin{eqnarray}
\label{wc}
&W_{cross}(e^-e^-;	p)&=\frac{e^4}{4(2\pi)^4}\int_{-\infty}^{+\infty} dqdkdldq_2dk_2dl_2\delta(p+q-k-l)\delta(p_2+q_2-k_2-l_2)
\delta(|p|+|q|-|k|-|l|)\frac{1}{(|p||q||k||l|)^2} \nonumber \\
&\times & \frac{1}{Z_{dir}Z_{ex}}\exp\Bigl(-\frac{k_2^2+l_2^2-Z_{dir}^2-Z_{ex}^2}{2eB}\Bigr)
\Bigl(\sqrt{\frac{\pi}{2}}\exp(-\frac{Z_{dir}|l_2|}{eB})
-\int_0^{\frac{Z_{dir}^2}{2eB}} dx \frac{\exp(-x-\frac{l_2^2Z_{dir}^2}{4xe^2B^2})}{\sqrt{2x}}\Bigr) \nonumber \\
&\times &\Bigl(\sqrt{\frac{\pi}{2}}\exp(-\frac{Z_{dir}|k_2|}{eB})
-\int_0^{\frac{Z_{dir}^2}{2eB}} dx \frac{\exp(-x-\frac{k_2^2Z_{dir}^2}{4xe^2B^2})}{\sqrt{2x}}\Bigr) \nonumber \\
& \times & \bigl((|p||k|+pk)(|q||l|+ql)-(|k|p+|p|k)(|l|q+l|q|)\bigr) \nonumber \\
&\times & \bigl((|p||l|+pl)(|q||k|+qk)-(|l|p+|p|l)(|k|q+k|q|)\bigr)\,\,f_0(q)(1-f_0(k))(1-f_0(l)),
\end{eqnarray}

and

\begin{eqnarray}
&W_{an}(e^+e^-;p)&=\frac{e^4}{(2\pi)^4} \int_{-\infty}^{+\infty} dqdkdldq_2dk_2dl_2\delta(p+q-k-l)\delta(p_2+q_2-k_2-l_2)
\delta(|p|+|q|-|k|-|l|) \nonumber \\
&\times & \frac{1}{(|p||q||k||l|)^2Z_{an}^2}\exp\Bigl(-\frac{(k_2+l_2)^2-Z_{an}^2}{eB}\Bigr)
\Bigl(\sqrt{\frac{\pi}{2}}\exp(-\frac{Z_{an}|k_2|}{eB})
-\int_0^{\frac{Z_{an}^2}{2eB}} dx \frac{\exp(-x-\frac{k_2^2Z_{an}^2}{4xe^2B^2})}{\sqrt{2x}}\Bigr)^2 \nonumber \\
&\times &(|p||q|-pq)^2(|k||l|-kl)^2\,\,f_0(q)(1-f_0(k))(1-f_0(l)),
\end{eqnarray}       
where $Z_{an}^2\equiv (k_2+l_2)^2-2(|p||q|-pq)$,

\begin{eqnarray}
&W_{dir}(e^+e^-;p)&=\frac{e^4}{4(2\pi)^4} \int_{-\infty}^{+\infty} dqdkdldq_2dk_2dl_2\delta(p+q-k-l)\delta(p_2+q_2-k_2-l_2)
\delta(|p|+|q|-|k|-|l|) \nonumber \\
&\times & \frac{1}{(|p||q||k||l|)^2Z_{dir}^2}\exp\Bigl(-\frac{k_2^2-Z_{dir}^2}{eB}\Bigr)
\Bigl(\sqrt{\frac{\pi}{2}}\exp(-\frac{Z_{dir}|k_2+l_2|}{eB})
-\int_0^{\frac{Z_{dir}^2}{2eB}} dx \frac{\exp(-x-\frac{(k_2+l_2)^2Z_{dir}^2}{4xe^2B^2})}{\sqrt{2x}}\Bigr)^2 \nonumber \\
&\times &\bigl((|p||k|+pk)(|q||l|+ql)-(|k|p+|p|k)(|l|q+l|q|)\bigr)^2\,\,f_0(q)(1-f_0(k))(1-f_0(l)),
\end{eqnarray} 

\begin{eqnarray}
&W_{ex}(e^+e^-;p)&=\frac{e^4}{4(2\pi)^4} \int_{-\infty}^{+\infty} dqdkdldq_2dk_2dl_2\delta(p+q-k-l)\delta(p_2+q_2-k_2-l_2)
\delta(|p|+|q|-|k|-|l|) \nonumber \\
&\times & \frac{1}{(|p||q||k||l|)^2Z_{ex}^2}\exp\Bigl(-\frac{l_2^2-Z_{ex}^2}{eB}\Bigr)
\Bigl(\sqrt{\frac{\pi}{2}}\exp(-\frac{Z_{ex}|k_2+l_2|}{eB})
-\int_0^{\frac{Z_{ex}^2}{2eB}} dx \frac{\exp(-x-\frac{(k_2+l_2)^2Z_{ex}^2}{4xe^2B^2})}{\sqrt{2x}}\Bigr)^2 \nonumber \\
&\times &\bigl((|p||l|+pl)(|q||k|+qk)-(|l|p+|p|l)(|k|q+k|q|)\bigr)^2\,\,f_0(q)(1-f_0(k))(1-f_0(l)),
\end{eqnarray}

\begin{eqnarray}
&W_{cross}(e^+e^-;p)&=\frac{e^4}{4(2\pi)^4} \int_{-\infty}^{+\infty} dqdkdldq_2dk_2dl_2\delta(p+q-k-l)\delta(p_2+q_2-k_2-l_2)
\delta(|p|+|q|-|k|-|l|)\frac{1}{(|p||q||k||l|)^2} \nonumber \\
&\times & \frac{1}{Z_{dir}Z_{ex}}\exp\Bigl(-\frac{k_2^2+l_2^2-Z_{dir}^2-Z_{ex}^2}{2eB}\Bigr)
\Bigl(\sqrt{\frac{\pi}{2}}\exp(-\frac{Z_{dir}|k_2+l_2|}{eB})
-\int_0^{\frac{Z_{dir}^2}{2eB}} dx \frac{\exp(-x-\frac{(k_2+l_2)^2Z_{dir}^2}{4xe^2B^2})}{\sqrt{2x}}\Bigr) \nonumber \\
& \times & \Bigl(\sqrt{\frac{\pi}{2}}\exp(-\frac{Z_{ex}|k_2+l_2|}{eB})
-\int_0^{\frac{Z_{ex}^2}{2eB}} dx \frac{\exp(-x-\frac{(k_2+l_2)^2Z_{ex}^2}{4xe^2B^2})}{\sqrt{2x}}\Bigr) \nonumber \\
&\times &\bigl((|p||k|+pk)(|q||l|+ql)-(|k|p+|p|k)(|l|q+l|q|)\bigr) \nonumber \\
&\times & \bigl((|p||l|+pl)(|q||k|+qk)-(|l|p+|p|l)(|k|q+k|q|)\bigr)\,\,f_0(q)(1-f_0(k))(1-f_0(l)),
\end{eqnarray}
where
the delta functions represent momentum and energy conservation.
All of the formulae involve the integration variables, $q,k,l,q_2,k_2$ and $l_2$.

Now we explain that the electric conductivity behaves as $\sigma \propto \beta^{-1}$
in the limit of $gB\beta^2 \to \infty$. 
In the integrals of $W$, we rescale all momenta such as e.g. $p=\beta^{-1}\bar{p}$.
Then, we find $W(p)=\beta^{-1}\bar{W}(\bar{p})$ in which dimensionless quantity $\bar{W}(\bar{p})$
depends on $\beta$ and $eB$ through the dimensionless quantity $eB\beta^2$.
Using the delta functions we can trivially perform the integration 
of the variables $\bar{q}$, $\bar{q}_2$ and $\bar{l}$.
Among remaining integrations of $\bar{k}$, $\bar{k}_2$ and $\bar{l}_2$, the integration of $\bar{k}$ is finite 
at $|\bar{k}|=\infty $ due to
the factor $f_0(\bar{q}=-\bar{p}+\bar{k}+\bar{l})\sim \exp(-|\bar{k}|)$ for $|\bar{k}| \to \infty$.
On the other hand, the integrations of $\bar{k}_2$ and $\bar{l}_2$ are 
finite at $|\bar{k}_2|=\infty $ or $|\bar{l}_2|=\infty $ owing to the factors
$\exp(-\bar{k}_2^2/eB\beta^2)$ or $\exp(-\bar{l}_2^2/eB\beta^2)$ and $1/Z_{\rm{dir}}^2\sim 1/\bar{k}_2^2$ 
or $1/Z_{\rm{ex}}^2\sim 1/\bar{l}_2^2$.
Hence, $W$ behaves such that $W(p)=\beta^{-1}\bar{W}(\bar{p})\propto \beta^{-1}eB\beta^2$
in the limit of $eB\beta^2 \to \infty$.
Therefore, $\sigma \propto \beta eB \int dp \tau(p) \propto \beta eB \int dp\, 1/W(p)
\propto \beta^{-1}$.

As we have stated before, 
the relaxation time $\tau\sim (eB\beta)^{-1}\sim (T/Q_s)1/Q_s$ is quite shorter
than the typical time scale $1/\sqrt{eE(t=0)}\sim Q_s^{-1}$ in the period of $T/Q_s \ll 1$.
Thus, we may think that the particles produced by Schwinger mechanism are immediately thermalized
at least in the early stage of the electric field decay.




\end{document}